# Utilization of a shallow underground laboratory for studies of the energy dependent CR solar modulation

N.Veselinović A. Dragić, M. Savić, D. Maletić, D. Joković, R. Banjanac, V. Udovičić,
*Institute of physics, University of Belgrade, Serbia*

The aim of the paper is to investigate possibility of utilizing a shallow underground laboratory for the study of energy dependent solar modulation process and to find an optimum detector configuration sensitive to primaries of widest possible energy range for a given site. The laboratory ought to be equipped with single muon detectors at ground level and underground as well as the underground detector array for registration of multi-muon events of different multiplicities. The response function of these detectors to primary cosmic-rays is determined from Monte Carlo simulation of muon generation and propagation through the atmosphere and soil, based on Corsika and GEANT4 simulation packages. The simulation predictions in terms of flux ratio, lateral distribution, response functions and energy dependencies are tested experimentally and feasibility of proposed setup in Belgrade underground laboratory is discussed. .

## 1. INTRODUCTION

Cosmic rays (CR) are energetic particles, arriving at the Earth from space after interaction with the heliosphere. The interaction of these, primary CRs, with the atmosphere leads to production of a cascade (shower) of secondary particles: hadrons, electrons and photons, muons, neutrinos. CR research has been undertaken at almost every location accessible to humans – from the outer space to deep underground [1].

At the low energy part of the spectrum, lower than 100 GeV, CRs are affected by the solar magnetic field. Modulation effects are energy dependent and have been studied extensively by the neutron monitors, sensitive up to about 10 GeV. Muon detectors at the ground level are sensitive to higher energy primaries [2], and the muons detected underground correspond to even higher energies. The possibility to further extend the sensitivity to higher energies with the detection of multi-muon events underground is the intriguing one. The idea was exploited with the EMMA underground array [3]. For a shallow underground laboratory, exceeding the energy region of solar modulation would open the possibility to study CR flux variations of galactic origin.

## 2. BELGRADE CR STATION

The Belgrade cosmic-ray station is situated at the Laboratory for Nuclear Physics at the Institute of Physics. Its geographic position is: latitude 44° 51' N and longitude 20° 23' E, altitude of 78 m a.s.l., with geomagnetic latitude 39° 32' N and geomagnetic vertical cut-off rigidity 5.3 GV. It is composed of two sections, the underground lab (UL) with useful area of 45 m$^2$, dug at the 12-meter shallow depth (equivalent to 25 m.w.e) and the ground level lab (GLL). At UL depth, practically, only the muonic component of the atmospheric shower is present.

The cosmic-ray muon measurements in Belgrade CR station are performed by means of the plastic scintillation detectors, placed both in the GLL and in the UL. With the previous set-ups, monitoring is continuous from 2002.

Measured cosmic-ray intensity data were thoroughly analysed, yielding some results on the variations of the cosmic-ray intensity [4,5,6].

Time series (pressure and temperature corrected) of these measurements can be accessed online at http://cosmic.ipb.ac.rs/muon_station/index.html.

In addition to single muon detectors, a small-scale test setup for multi-muon events is installed underground. It consists from three scintillators: one large detector (100cm x 100cm x 5cm) and two small detectors (50cm x 23cm x 5cm) which are placed horizontally on their largest sides. Their mutual position is adaptable. The data acquisition system is based on fast 4-channel flash analog-to-digital converters (FADC), made by CAEN (type N1728B), with 100 MHz sampling frequency. The events are recorded in the list mode. For each event from every input channel the timing and amplitude are saved, together with auxiliary information such as the result of pile-up inspection routine. From this list a time series of single or coincident events could be constructed. The experimental set-up is sketched in Figure 1.

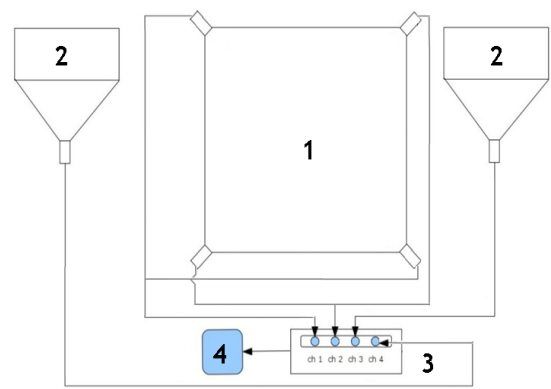

Figure 1 : Sketch of the experimental set-up for the cosmic-ray measurements:
1) Large scintillation detector, 2) small scintillation detectors, 3) flash analog-to-digital Converter (FADC), 4) experiment control and data storage computer

With simultaneous operation of several detector systems, as described, a single facility with the same rigidity cut-off would be used for investigation of solar modulation at different energies. Further integration with the Neutron Monitors data would be beneficial [7, 8, 9].

## 3. SIMULATION DETAILS AND RESULTS

Simulation of the CR shower dynamics up to the doorstep of GLL and UL has been done using Monte Carlo simulation packages CORSIKA and Geant4 [10, 11]. The cosmic-ray muon spatial and momentum distribution at 78m a.s.l. is of our interest. The output of CORSIKA at ground level is used as the input for Geant4 based simulation of particle transport through the soil and simulation of detector response. For this purpose soil analysis is done beforehand. The mean density is found to be (2.0±0.1) g/cm3 and soil type is loess with the assumed composition of Al2O3 20%, CaO 10% and SiO2 70%. For the simulation of underground detector system only those muons with energy sufficient enough to survive passage through soil are taken into consideration (Figure 2).

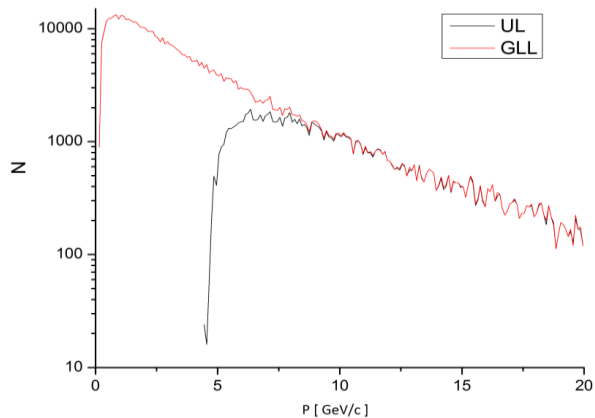

Figure 2: Surface momentum distribution for muons at GLL and muons reaching UL at Belgrade CR station based on GEANT4 and CORSIKA

At lower energies, protons make ~85% of CR, so primary particles used in the simulation were protons. The number of muons reaching UL is not linearly proportional to energy of the primary particle, especially for energies lower than 200 GeV which is energy range of interest, as showed in Figure 3. This correspond to similar work done elsewhere [12]. Probability that a registered event corresponds to a primary particle of certain energy is inferred from the simulation for every detection system:
- Single muon detector at ground level
- Single muon detector underground
- Two-fold muon coincidences underground
- Muon coincidences of higher multiplicity

For these response functions, simulation use 23 million primary protons with energy range from 5 GeV to $10^{16}$ eV with zenith angle between (0°, 70°) and with power law energy spectrum with the exponent -2.7.

Shift toward higher energies is evident for transition from GL to UL and to the events of higher multiplicities.

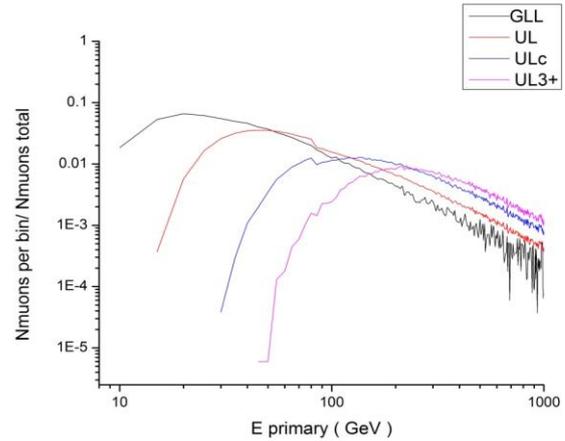

Figure 3: Differential response functions of muon detectors in GLL and UL based on simulation for: single muons at ground level (GLL), underground level (UL), coincident muons at underground level (ULc) and triple and higher multiplicity coincident muons at underground level (UL3+) normalized to total number of muons respectively.

For all relevant quantities of the muon flux is given at Table 1. Equivalent depth was found using ratios of integral fluxes of muons at different shallow depth [13].

Table 1: Properties of the flux of the primary particles at Belgrade CR station based on simulation for: ground level (GLL), underground level (UL), coincident muons at underground level (ULc) and triple and more coincident muons at underground level (UL3+).

| Primary protons | GLL | UL | ULc | UL3+ |
|---|---|---|---|---|
| Energy cut-off | 5 GeV | 12.3±0.7 GeV | 30±4 GeV | 55±14 GeV |
| Equivalent depth | 0 m.w.e. GLL Belgrade | 25 m.w.e. | 40 m.w.e. | 66 m.w.e. |
| Peak energy | 20 GeV | 45 GeV | 125 GeV | 200 GeV |
| Median energy | 62 GeV | 122±5 GeV | 296 ± 8 GeV | 458±18 GeV |

Cut-off energy at the ground level is due to geomagnetic cut-off rigidity at Belgrade CR station. For the underground level, the 25 m.w.e. of soil overburden is the cause of the higher cut-off underground. All the relevant quantities: cut-off, peak and median energies are higher underground and for the events with higher multiplicity. This, in principle, creates a possibility to investigate the CR flux and its variations at different energies of primaries, exceeding the energies relevant to neutron monitors, the most frequently used instrument for study of the low energy side of the CR spectrum. This vindicates the aim of the simulation to investigate possibility of utilizing a shallow underground laboratory for the study of energy dependent solar modulation process and to find an optimum detector configuration sensitive to primaries of different energy range for a given site.

## 4. DISCUSION ON FEASIBILITY

It is needed, however, to address the questions of reliability of simulation. On the graph 3, the discontinuity at energy of 80 GeV of primary protons is visible, especially muon in UL and muons in UL in coincidence. CORSIKA, by default, uses GHEISHA 2002d particle generator to calculate the elastic and inelastic cross-sections of hadrons below 80 GeV in air and their interaction and particle collisions and for higher energies QGSJET 01C routine is used. Also it is important to know whether sufficient statistics of multi-muon events could be achieved in the limited laboratory space. For this purpose, the flux of single muons is measured at ground level and underground, the rate of double coincidences as a function of detector distance is simulated also. In addition, the rates of double and triple coincidences are also measured for several detector arrangements.

The muon flux is calculated, from simulation, by finding ratio between the number of muons reaching depth of UL ( for single and muons in coincidence) and numbers of muons generated from CORSIKA at the surface and multiplying by experimentally measured value of integrated muon surface flux which is 137(6) muons per $m^2s$ [14]. The experimental value of integrated flux, compared with number of muons from simulation, is also used to find physical time needed to generate same number of muons at the site as the simulation.

Absolute muon fluxes measured at the site for surface and shallow underground is well reproduced by the simulation (Table 2).

Table 2: Ratio of muon fluxes at Belgrade CR station based on measurements and simulation for: ground level (GLL), underground level (UL), coincident muons at underground level (ULc) and triple and more coincident muons at underground level (UL3+)

| Muon flux ratio | Measured GLL/UL | Simulation GLL/UL | Simulation UL/ULc | Simulation UL/UL3+ |
|---|---|---|---|---|
| | 3.17(8) | 3.06(3) | 1.86(4) | 2.68(6) |

Recently with new detector arrangement, the scintillators in Belgrade CR station measured coincident events and triple coincident events at two distances of the detectors: 1.5m and 6m, in UL part of the laboratory. Number of coincidences per unit area of the detector, based on simulation for these distances is 80 and 66 muon coincidences per $m^2$ per day respectively. Experimental values are higher for closer ( ~350 coincidences a day ) and ~60 coincidences per day for farther arrangement. The ratio of single/coincidence events underground is well reproduced for greater distance of the detector. At shorter distances the measured ratio is higher than predicted by simulation, further study will show is it due to contribution from local EM showers and knock-on electrons. Numbers of measured triple coincidences at same distances are the order of magnitude smaller.

When upgraded, the detector arrangements will cover the whole area of the UL with muon detectors it should provide, based on the simulation, approximately 61k coincidences per day thus allowing to observe ~ 1.2% fluctuation of the CR flux with 3σ certainty originated from Solar modulation ( e.g. Forbush decreases) thus allowing possibility to study solar modulation on three different energy ranges of the primary particles and at higher energies then regular energies detected with NM. To prevent miss-identification of muons, additional methods of sorting muons is needed (lead shielding, hodoscopes...) or to measure only coincidences that occur on reliable distances between detectors, larger then 6m, allowing observation of higher fluctuations (~2.5%) with same certainty.

In principle, larger shallow depth laboratories [15] can be used to investigate solar modulation and extreme solar events on different energies of primary particles, using rate of detected muons on different detectors in coincidence but present small detection area at Belgrade CR station can also give some valuable insight.

## 5. CONCLUSION

The possibility of utilizing a shallow underground laboratory for the study of energy dependent solar modulation of CR is investigated, by means of computer simulation based on CORSIKA and GEANT packages, combined with the experiment. On the experimental part, the muon flux is measured at ground level and underground at the depth of 25 mwe. In the present feasibility study, the flexible test setup for detection of multiple muons is installed underground in an attempt to achieve sensitivity to higher energy primaries. The rates of double and triple coincidences are measured for several detector distances. The simulation revealed the response functions of each experimental setup. The experimental fluxes are compared with those arising from simulation (Table 2). For single muons, the experimental ratio of fluxes GLL/UL agrees with the simulated one. The experimental ratio of single/coincident events underground is well reproduced by simulation if the detector distance is greater than 6m. At shorter distances the measured ratio is higher than predicted by simulation, mainly due to contribution from local EM showers and knock-on electrons. When upgraded, the detector arrangements will cover almost the entire area of the UL with muons detectors resulting in expected approximately 61k coincidence per day. One day of measurements will be sufficient to observe ~ 1.2% fluctuation of the flux at $3\sigma$ significance for CRs with several hundred GeV of energy. Together with the single muon measurements at GLL and UL we will have simultaneous measurements centered on three different energies, under the same atmospheric and geomagnetic conditions. Any difference in time series behavior could be attributed to energy dependent response to the forcing. The rate of triple coincidences is too low to be effectively exploited in our conditions.

## Acknowledgments

The present work was funded by the Ministry of Education and Science of the Republic of Serbia, under the Project No. 171002. .